\documentclass[aps,prx,twocolumn,longbibliography,superscriptaddress,preprintnumbers,10pt]{revtex4-2}
\usepackage[colorlinks,bookmarks=true,citecolor=blue,linkcolor=blue,urlcolor=blue]{hyperref}
\usepackage{dcolumn,graphicx,amsfonts,amsthm,bm,color,appendix,float,tabularx}
\usepackage{braket}
\usepackage[normalem]{ulem}
\usepackage[version=4]{mhchem}
\usepackage{siunitx}
\usepackage{CJKutf8}
\usepackage{multirow}
\usepackage{booktabs}
\usepackage{soul}
\usepackage{graphicx}
\usepackage{braket}
\usepackage{CJKutf8}
\usepackage{placeins}
\usepackage[capitalize]{cleveref} 

\usepackage[dvipsnames]{xcolor}
\definecolor{pal0}{rgb}{0.8941, 0.102 , 0.1098}
\definecolor{pal1}{rgb}{0.2157, 0.4941, 0.7216}
\definecolor{pal2}{rgb}{0.302 , 0.6863, 0.2902}
\definecolor{pal3}{rgb}{0.5961, 0.3059, 0.6392}
\definecolor{pal4}{rgb}{1.    , 0.498 , 0.    }
\definecolor{pal5}{rgb}{0.83, 0.69, 0.22}

\newcommand{\COMMENTED}[1]{}

\newcommand{\n}[1]{\left| #1 \right|}
\newcommand{\bs}{\boldsymbol}

\let\oldv\v
\renewcommand{\v}[1]{\boldsymbol{#1}}

\def \R {\mathbf{R}}
\def \r {\boldsymbol{r}}
\def \k {\boldsymbol{k}}

\newcommand{\PRLsection}[1]{\section{#1}}

\begin{document}
\title{Quantum Geometry Driven Crystallization:\\ A Neural-Network Variational Monte Carlo Study}

\author{Agnes~Valenti}
\thanks{A.V. and Y.V. contributed equally to this work.}
\affiliation{Center for Computational Quantum Physics, Flatiron Institute, New York, NY, 10010, USA}
\author{Yaar~Vituri} 
\thanks{A.V. and Y.V. contributed equally to this work.}
\affiliation{Department of Condensed Matter Physics, Weizmann Institute of Science, Rehovot 76100, Israel}
\author{Yubo~Yang~(\begin{CJK*}{UTF8}{gbsn}杨煜波\end{CJK*})}
\affiliation{Department of Physics and Astronomy, Hofstra University, Hempstead, New York 11549, USA}
\author{Daniel~E.~Parker}
\affiliation{Department of Physics, University of California at San Diego, La Jolla, California 92093, USA}
\author{Tomohiro Soejima (\begin{CJK*}{UTF8}{bsmi}副島智大\end{CJK*})}
\affiliation{Department of Physics, Harvard University, Cambridge, MA 02138, USA}
\author{Junkai~Dong (\begin{CJK*}{UTF8}{bsmi}董焌\end{CJK*}\begin{CJK*}{UTF8}{gbsn}锴\end{CJK*})}\affiliation{Department of Physics, Harvard University, Cambridge, MA 02138, USA}
\author{Miguel A. Morales}
\affiliation{Center for Computational Quantum Physics, Flatiron Institute, New York, NY, 10010, USA}
\author{Ashvin~Vishwanath}
\affiliation{Department of Physics, Harvard University, Cambridge, MA 02138, USA}
\author{Erez~Berg}
\affiliation{Department of Condensed Matter Physics, Weizmann Institute of Science, Rehovot 76100, Israel}
\author{Shiwei~Zhang}
\affiliation{Center for Computational Quantum Physics, Flatiron Institute, New York, NY, 10010, USA}

\date{\today}

\begin{abstract}
Wigner crystals are a paradigmatic form of interaction driven electronic order. A key open question is how Berry curvature and, more generally, quantum geometry reshape crystallization. The discovery of two-dimensional materials with relatively flat bands and pronounced Berry curvature has added fresh urgency to this question.
Recent mean-field studies have proposed a topological variant of the Wigner crystal, the anomalous Hall crystal (AHC), with non-zero Chern number. However it remains unclear whether the AHC survives 
beyond the mean-field approximation.
Here, we map out the ground-state phase diagram of the $\lambda$-jellium model --- 
a simple model whose interaction strength and Berry curvature are independently tunable ---
using state-of-the-art neural-network variational Monte Carlo.
The AHC is found to remain stable against quantum fluctuations.
Surprisingly, quantum geometric effects are found to dramatically enhance crystallization. Both the AHC and  the standard Wigner Crystal are stabilized at densities up to an order of magnitude above the critical density in the absence of quantum geometry, yet still significantly below the threshold predicted by mean-field theory.
These striking results highlight the rich interplay between quantum fluctuations, quantum geometry, and crystallization, providing concrete guidance for  experiments and enabling future explorations of fractionalized crystals and chiral superconductors. 
\end{abstract}

\maketitle

\PRLsection{Introduction}
Under strong interactions, electrons can spontaneously break continuous translational symmetry, forming electron crystals with an emergent lattice structure~\cite{wigner_interaction_1934}.
The paradigmatic model for studying this effect is jellium:  a Galilean invariant system of electrons interacting via the Coulomb interaction in a rigid neutralizing background.  In this system, the physics is governed entirely by the electron density, which determines the ratio of potential to kinetic energy, denoted $r_s$. For any dimension $d>1$, robust scaling arguments yield
a Fermi liquid at weak interactions (high densities or low $r_s$) and a Wigner crystal (WC) at the strongly interacting (low density) limit $r_s \to \infty$, with the critical value of $r_s$ dependent on details~\cite{Giuliani_Vignale_2005}. 
For a long time, Wigner crystals in electronic systems remained elusive, with the first indirect evidence appearing on the surface of liquid helium~\cite{grimes_evidence_1979}, and later in semiconductors under large magnetic fields~\cite{lozovik_crystallization_1975,andrei_observation_1988,santos_observation_1992}. Recent advances in van der Waals materials have reignited interest in this century-old problem, providing new experimental platforms that offer new evidence of Wigner crystallization~\cite{smolenski_signatures_2021,tsui_direct_2023, zhou_bilayer_2021, 
xiang_quantum_2024,sung2025electronic}.

At the same time, a plethora of novel two-dimensional electron platforms discovered in recent years have stressed the importance of a crucial element which is not present in jellium: quantum geometry and topology. This raises an immediate question: how does non-trivial quantum geometry affect electron crystallization?
A number of studies have uncovered new phenomenology arising in the presence of non-trivial quantum geometry \cite{AHC_Yahui,AHC_Senthil,AHC1,Zeng_sliding,joy2023wignercrystallizationbernalbilayer, desrochers2025elasticresponseinstabilitiesanomalous, AHC4, tan2025ideallimitrhombohedralgraphene, Zeng_sublattice_structure, Tan_parent_berry, Tan_FAHC, AHC2, AHC3, bernevig2025berrytrashcanmodelinteracting, patri2025familymultilayergraphenesuperconductors}, such as modified dynamics in the absence of Galilean invariance~\cite{Zeng_sliding}, new geometries and types of Wigner crystals~\cite{desrochers2025elasticresponseinstabilitiesanomalous,joy2023wignercrystallizationbernalbilayer,AHC3}, and unconventional phonon dynamics~\cite{AHC4,tan2025ideallimitrhombohedralgraphene}. A particularly striking consequence of nontrivial quantum geometry, inspired by recent experiments in rhombohedral graphene~\cite{lu2024fractional}, is the proposal~\cite{AHC1, AHC_Yahui} of \textit{anomalous Hall crystals} (AHCs): electron crystals which have a nonzero Chern number.
These topological crystals are stabilized by the existence of nontrivial parent band geometry~\cite{Zeng_sublattice_structure,Tan_parent_berry,Zhihuan_Stability,AHC2,AHC3,bernevig2025berrytrashcanmodelinteracting}.
In contrast to previously proposed Hall crystals
~\cite{tesanovic_hall_1989,kivelson_cooperative_1986, halperin_compatibility_1986, kivelson_cooperative_1987, murthy2000}, whose topology derives from Landau levels, the AHC requires no magnetic field, and its topology and crystallization emerge on equal footing.

To tackle the role of nontrivial quantum geometry, some of us have proposed a minimal topological extension of the jellium model (``$\lambda$-jellium'') in Ref.~\cite{AHC3}. 
This model retains the quadratic band dispersion and Coulomb interaction, whose strength is quantified by $r_s$, and introduces an additional parameter $\lambda$ that adjusts the concentration of Berry curvature.
At the mean-field level, the ground state transitions from a Wigner crystal to an AHC at around $\lambda \approx 1/2$.

Strong correlations limit the validity of mean-field approaches.
This is manifest in the paradigmatic jellium model, where the 2D liquid-crystal transition is found within mean-field approximation to occur at $r_s^\star \approx 2$. This value is off by more than an order of magnitude from the realistic value: critical $r_s^\star \gtrsim 25$, found in quantum Monte Carlo (QMC) calculations ~\cite{tanatar1989ground,Drummond_Phase_Diagram}.  
Thus, it is of immediate relevance to ask: Are AHCs stable under quantum fluctuations beyond mean-field? 
More generally, how do Berry curvature and quantum geometry affect the crystallization of electrons?

In this paper,
we address these questions by a systematic study of the 
$\lambda$-jellium model,
employing state-of-the-art variational Monte Carlo (VMC) with neural quantum states (NQS).
Such 
neural-network-based VMC
approaches have become
a powerful tool
for  many-body problems defined in continuum space
~\cite{pescia2024message, smith2024ground,luo2024simulating,teng2025solving, luo2025solving, geier2025attention, teng2025solving, li2025emergent, li2025attention}.
For example, (MP)$^2$NQS~\cite{smith2024ground} - an NQS ansatz
based on a Slater-Jastrow-Backflow wave-function with message-passing graph neural-network components~\cite{Pescia2024MPNQS},
have achieved significantly lower energies 
than state-of-the-art QMC 
in two-dimensional jellium~\cite{smith2024ground}. 
Here, we leverage the (MP)$^2$NQS ansatz and generalize it to allow non-trivial spinor structures, which enables its application to the $\lambda$-jellium model as well as many other systems.

\begin{figure}[t]
    \centering
    \includegraphics[width=0.5\textwidth]{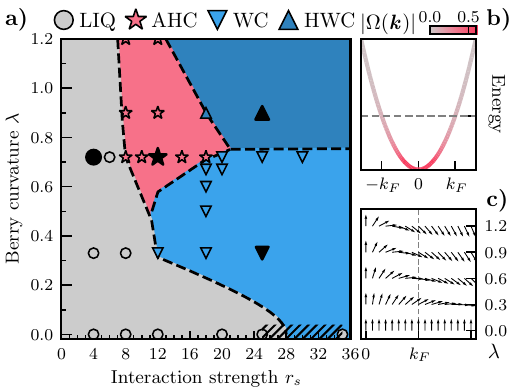}
    \caption{
    \textbf{(a)} 
    Phase diagram of the $\lambda$-jellium model, computed by neural-network VMC.
    The ground state is shown as a function of the two independent parameters, the 
    interaction strength and quantum geometry. 
    At $\lambda=0$,
    the Wigner crystal (WC) competes with a Fermi liquid (LIQ), with a crystallization transition at $r_s^\star \gtrsim 25$. The dashed region indicates a region in which the ground state starts showing first crystalline features (for more details, see App.~\ref{app:boundaries}).
    As topology is added with $\lambda$,
    the critical $r_s^\star$ for WC transition decreases significantly. At sufficient $\lambda$, 
    a ``halo Wigner crystal" (HWC) appears at large $r_s$, and an anomalous Hall crystal (AHC) is stabilized at intermediate $r_s$. 
    Symbols indicate numerical datapoints in each phase, with filled symbols indicating data points for which we plot the observables in Fig~\ref{fig:phase_identification}. Dashed line indicate interpolated phase boundaries (see App.~\ref{app:boundaries}).
    \textbf{(b)} Quadratic dispersion of $\lambda$-jellium's lower band at $\lambda=0.72$. The line color shows the Berry curvature, which is concentrated at $\k=0$.
    \textbf{(c)} The skyrmionic spinor texture of the lower band of $\lambda$-jellium along $k_x$ (taking $k_y=0$). The skyrmion core shrinks with increasing $\lambda$, concentrating the Berry curvature.
    }
    \label{fig:phase_diagram}
\end{figure}

We find that the anomalous Hall crystal phase is stable in the presence of quantum fluctuations.
Remarkably, incorporating quantum geometry reduces the crystallization threshold greatly, from $r_s^\star \gtrsim 25$ to as low as $r_s^\star \approx 8$ for transition into the AHC, and 
r$_s\approx 12$ for that into the WC.
These findings are unified in Fig.~\ref{fig:phase_diagram}, where we show delicate competitions among the AHC, Fermi liquid, WC, and ``halo Wigner crystal'' (HWC) phases, all of which are defined below. 

In the following, we introduce our model and numerical methods, then describe the many-body phase diagram of $\lambda$-jellium, and conclude with experimental implications. 

\PRLsection{$\lambda$-jellium Model}

The $\lambda$-jellium model~\cite{AHC3} has a two-spinor Hamiltonian
\begin{equation}
\label{eq:ham}
	\hat{H}
    = \Delta \sum_{i=1}^N \begin{bmatrix}
        -\lambda^2 \nabla_i^2 & i\lambda \partial_i\\
        i\lambda \bar{\partial}_i & 1
    \end{bmatrix}
    - 
	\sum_{i=1}^N \hat{I}_2 \frac{\nabla_i^2}{r_s^2} + \frac{2}{r_s} \sum_{i<j}^N \frac{1}{\n{\v{r}_i - \v{r}_j}},
\end{equation}
where $\hat{I}_2$ is the identity in spinor space, $\partial = \partial_x - i \partial_y, \nabla^2 = \partial_x^2 + \partial_y^2$, and the interaction is spinor isotropic.
The first term, responsible for generating topology, was also written down in Refs.~\cite{bernevig2006quantum,Tan_parent_berry,hu2018fractional} up to a diagonal piece.
In the absence of this topologically non-trivial first term, this model reduces to the standard jellium model.
The  dimensionless parameter $r_s$ is defined in the same manner 
as in jellium, and 
a uniform neutralizing charge background 
is assumed.
We measure 
all energies in units of \si{Ry} and lengths in units $r_s a_B$, 
where $a_B = 4\pi \epsilon_0 \hbar/me^2$ is  the Bohr radius. 

Eq.~\eqref{eq:ham} has continuous translation symmetry and $U(1)$ rotation symmetry $\hat{R}_\theta \phi(\v{r}) = M_{\theta} \phi(R_{-\theta} \v{r})$ where $M_{\theta} = \mathrm{diag}[1, e^{i\theta}]$ acts in spinor space.
The lower band, which controls the low energy physics, has quadratic dispersion $\epsilon_{1,\v{k}}=\n{\v{k}}^2/r_s^2$ and wavefunction
$\phi_{\v{k}}(\v{r})=~ \begin{bmatrix} 1, & \lambda (k_x+i k_y) \end{bmatrix}^T e^{i\v{k}\cdot\v{r}}$ (see also Refs.~\cite{hu2018fractional, Tan_parent_berry}). In the liquid phase, the Fermi wave vector has length $k_{F}=2$. The spinor part has a skyrmionic texture~(Fig.~\ref{fig:phase_diagram}c), that produces concentrated Berry curvature
$\Omega(\v{k}) = - 2 \lambda^2(\lambda^2 \n{\v{k}}^2 + 1)^{-2}$, which seeds crystals with a negative Chern number~\cite{AHC2, Tan_parent_berry, Zhihuan_Stability}. We set $\Delta = 2/r_s$ such that the second band has parametrically larger kinetic energy $\epsilon_{2,\v{k}} \ge 2/r_s$, suppressing its role in the ground state (see App.~\ref{app:numerical_robustness}, Fig.~\ref{fig:band_mixing}).
Thus the low energy sector of Eq.~\eqref{eq:ham} has a quadratic band minimum with Coulomb interaction strength $r_s$ and the independent parameter $\lambda$ controls its Berry curvature, reducing to polarized jellium when $\lambda \to 0$.

\PRLsection{Method} 
In our neural-network VMC, 
we parametrize the ground state using a Slater-Jastrow-backflow (SJB) ansatz,
(MP)$^2$NQS~\cite{smith2024ground}, that represents electrons with an internal spinor structure
\begin{equation}
    \Psi\left( \R,S \right) = e^{-U\left( \R,S \right)} \det\left[ \phi_{j} \left( \bs{r}_i+\mathcal{N}_i(\R),\sigma_i \right) \right]~,
    \label{eq:wf-ansatz}
\end{equation}
with $\R=(\r_1,\dots,\r_N)$ the many-body coordinate and $S=(\sigma_1,\dots,\sigma_N)$ the many-body spinor index in $\hat{z}$ basis ($\sigma_i=\pm1$). Each orbital $\phi_{j}$ is a linear combination of plane-waves with optimizable coefficients:
\begin{equation}
    \phi_j(\bs{r}, \sigma) = \sum\limits_{\k} c^{(\sigma)}_{j\k} e^{i\k\cdot\r}.
\end{equation}
In Eq.~(\ref{eq:wf-ansatz}), the orbitals are evaluated at the backflow transformed 
coordinate $\tilde{\r}_i =\r_i + \mathcal{N}_i(\R)$, where 
$\mathcal{N}_i$ denotes a many-body message-passing neural network (MPNN).
The Jastrow factor is given by a sum of two contributions
\begin{equation}
\label{eq:jastrow}
    U\left( \R,S \right) = U_2\left( \R,S \right) + U_{N.N.}( \R),
\end{equation}
where $U_2$ is a two-body spinor-dependent Jastrow: $U_2=\sum_{i\neq j}u_{\sigma_i,\sigma_j}(|\r_i-\r_j|)$, and each $u_{\sigma,\sigma'}$ is encoded as a B-spline. The second term, $U_{N.N.}$, is a complex function derived by applying a multilayer perceptron (MLP) to the backflow transformed coordinates.
The strategic placement of the MPNN and
MLP into the SJB ansatz, which 
by itself already gives an excellent 
description of jellium, makes a powerful approach. It is flexible and accurate
while remaining sufficiently compact 
to be optimized in large systems.

The orbital coefficients, the weights of the two-body B-spline functions, the backflow MPNN parameters and the many-body Jastrow MLP are optimized using minimum-step stochastic reconfiguration~\cite{sorella2001generalized,chen2024empowering} to minimize the variational energy $E=\frac{\braket{\Psi|\hat{H}|\Psi}}{\braket{\Psi|\Psi}}$. 
We choose simulation cells and particle numbers commensurate with a triangular lattice, and optimize two sets of wave-functions: On one set, we impose discrete lattice translational invariance on the orbitals, thereby reaching high-accuracy crystalline states. On the second set, we lift this restriction to robustly optimize the liquid states. All simulations are performed on $N_e=36$ electrons unless stated otherwise. 
We show the robustness of our results to finite size effects, twisted boundary conditions, and more expressive ans\"atze in  App.~\ref{app:numerical_robustness}.
We provide more details on the wave-function and optimization in App.~\ref{App:ansatz}.

\PRLsection{Phase Diagram}

The main result of this paper is the phase diagram of $\lambda$-jellium within NQS-VMC, shown in Fig.~\ref{fig:phase_diagram}a.
Each symbol indicates the ground-state phase at the given $(r_s,\lambda)$ determined by a 
full set of calculations. 
Four competing phases are identified: a Fermi liquid (LIQ), and three crystalline phases. These are: (1) a Wigner crystal (WC), with Chern number $C=0$; (2) an anomalous Hall crystal (AHC)~\cite{AHC_Yahui, AHC1}, with Chern number $C=-1$; and (3) a ``halo" Wigner crystal (HWC)~\cite{joy2023wignercrystallizationbernalbilayer, AHC3}, also with $C=0$. The HWC is adiabatically connected to a Wigner crystal composed of Wannier orbitals with angular momentum $L_z^{\mathrm{Wan}} = 1$, which manifests as a ``halo" pattern in its momentum distribution $n(\v{k})$. 
Computational details, including the interpolation procedure to determine phase boundaries, are given in App.~\ref{App:ansatz} and \ref{app:boundaries}.

\begin{figure}
    \centering
\includegraphics[width=0.5\textwidth]{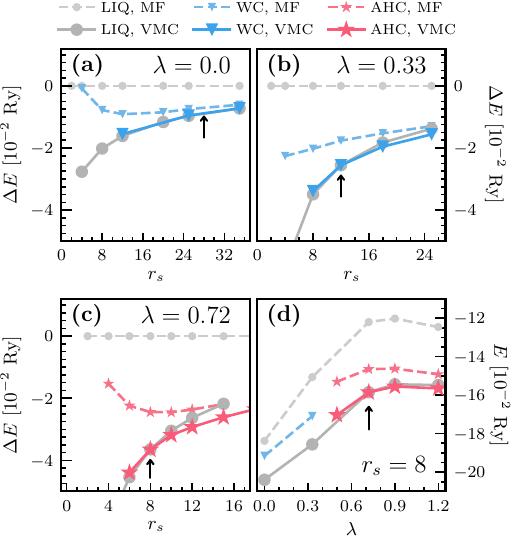}
    \caption{
    The delicate balance 
    of the energetics in different phases, and the effect from correlation and quantum geometry.
    \textbf{(a)} In the usual 
   2D jellium (polarized), the energy of the WC phase becomes lower than the LIQ phase at 
   $r_s^\star \gtrsim 25$ (see Appendix~\ref{app:boundaries} for more details on the $\lambda=0$ transition, here we mark the literature value $r_s^\star \approx 28$ \cite{Drummond_Phase_Diagram} with a black arrow). Mean-field yields a transition 
   at a much smaller $r_s$, indicated 
   by the crossing of the dashed lines around $r_s^\star \approx 2$. Correlation energy is much larger in the LIQ phase.
   \textbf{(b-c)} Quantum geometry lowers the critical value of the liquid-crystal transition.
   These have the same setup as in \textbf{(a)}, but are
   for LIQ and WC at $\lambda=0.33$ and LIQ and AHC at $\lambda=0.72$. The black arrows mark the VMC transitions estimated via our simulations.
   \textbf{(d)}
   The total energies are given as a function of the Berry curvature concentration, 
for a fixed $r_s=8$.
    }
    \label{fig:energetics}
\end{figure}

We use energetics to identify the ground-state phase at each $(r_s,\lambda)$, 
as illustrated in Fig.~\ref{fig:energetics}, followed by 
detailed studies of the 
``fingerprints" of each phase 
to characterize them, as discussed in the next section.  
At $\lambda=0$, where the model reproduces polarized jellium,  the Fermi liquid transitions to a WC at $r_s^\star \gtrsim 25$ (see Fig.~\ref{fig:energetics}(a)),
matching existing literature up to finite size effects~\cite{Drummond_Phase_Diagram}. Adding quantum geometry shifts the transition to dramatically higher densities, i.e., smaller $r_s$. 
Fig.~\ref{fig:energetics}(b) shows that at $\lambda=0.33$, crystallization occurs at $r_s^\star \approx 12$ --- more than four times the density required in polarized jellium. 
As we further increase the concentration of Berry curvature, the crystalline phase
becomes topological. Fig.~\ref{fig:energetics}(c) shows that at $\lambda=0.72$, the liquid transitions to an AHC at $r_s^{\star}\approx 8$.
The AHC phase persists up to $r_s^\star \approx 18$, whereupon it transitions into a WC (upon further increasing $\lambda$, this WC  transitions into a HWC). 
In comparison to mean-field 
 calculations (see Fig.~\ref{fig:energetics}), overall our VMC calculation finds drastic phase-boundary shifts and a much more delicate energetic balance.

A striking feature emerging from the many-body solution is the significant decrease of 
critical $r_s^\star$ for crystallization with 
$\lambda$. 
As Fig.~\ref{fig:energetics}(d) shows, for fixed $r_s$, 
the energies increase when $\lambda$ is turned on, 
originating from the penalty on the exchange energy due to the non-trivial spinor texture at $\lambda>0$.
As we detail in App.~\ref{app:energetics_details}, this penalizes the liquid more strongly than the crystal, increasing the difference in interaction energy by more than an order of magnitude. We find that this is due to the short-range part of the interaction: the exchange-correlation hole is deformed with increasing $\lambda$ for both phases, but the deformation is larger in the liquid.
Simultaneously, the energetic competition scale~\footnote{We define the energetic competition scale as $\frac{d (E^{\rm{LIQ}} - E^{\rm{crys}})}{dr_s}$ at $r_s^\star$.} between crystal and liquid is increased substantially 
--- to ${\sim}10^{-3}$\si{Ry} 
from 
the $\lambda=0$ scale of ${\sim}10^{-4}$\si{Ry}. (It is however still 
much smaller than the mean-field scale of 
${\sim}10^{-2}$\si{Ry}.)

Why does quantum geometry promote crystallization?
At finite $\lambda$, the states are no longer spinor-polarized, allowing electrons with orthogonal spinors to occupy the same point in space.
We show in App.~\ref{app:energetics_details} that the metals are less able to keep the electrons apart than crystalline states, manifesting in a larger probability to be at the same point and, consequently, a larger penalty from the interaction energy. The delicate energetic competition at $\lambda=0$ is therefore tilted in favor of the crystal at $\lambda >0$, dramatically shifting the phase boundary.

\begin{figure*}
    \centering
    \includegraphics[width=\linewidth]{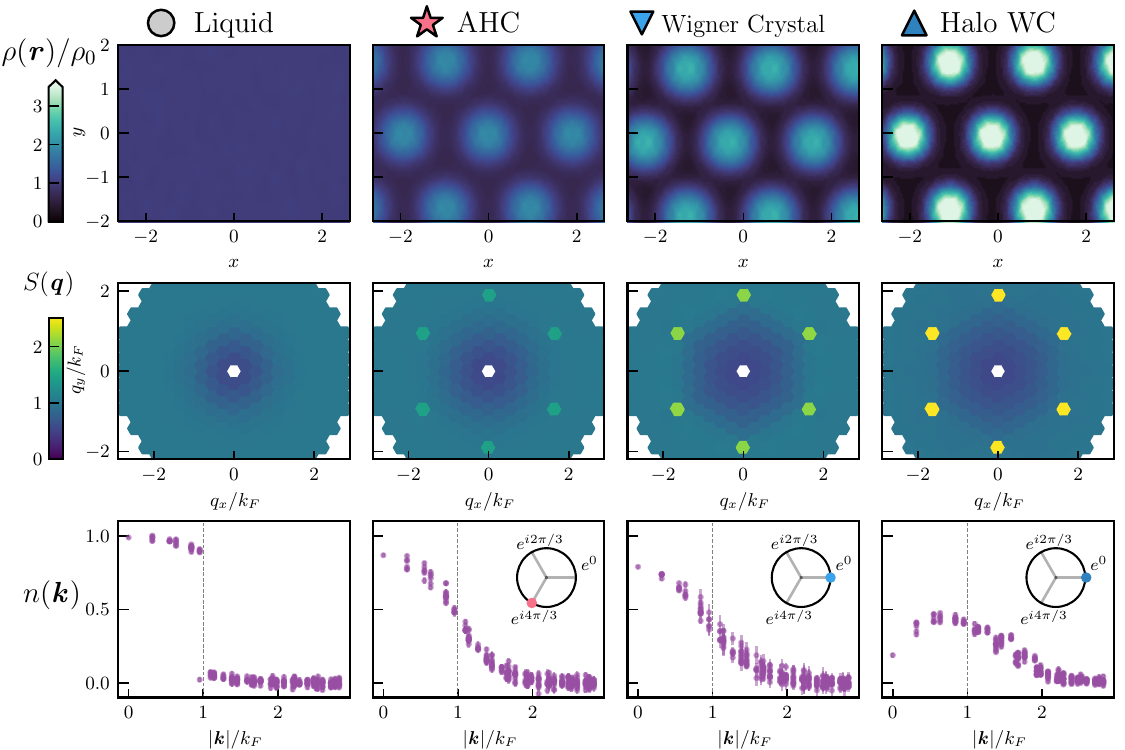}
    \caption{
    Properties of the different 
    phases from the variational NQS ground states.
    \textbf{Columns} correspond to filled symbols in Fig.~\ref{fig:phase_diagram}, at parameters $(r_s,\lambda)$ of $(4,0.72)$ LIQ, $(12,0.72)$ AHC, $(25,0.33)$ WC, and $(25,0.9)$ HWC. 
    \textbf{Row 1}: charge density $\rho(\r)$ of each state, normalized by the average charge density $\rho_0$. 
    \textbf{Row 2}: static structure factor $S(\v{q})$. The six sharp Bragg peaks  indicate crystallization to a triangular lattice with one electron per unit cell. \textbf{Row 3}: momentum space occupations $n(\v{k})$ for each phase versus unrestricted momentum $\n{\v{k}}/k_F$. The liquid has a sharp Fermi surface (dotted line at $k_F$). The HWC has small maximal occupations in a ``halo" peaked near $0.6k_F$ in this case, and a minium at $\n{\v{k}}=0$. Error bars show statistical error.
    \textbf{Insets:} Monte Carlo estimates of $\braket{\Psi|C_3|\Psi} = e^{i2\pi C/3}$ where $C$ is the Chern number (see text). The WC and HWC have $C=0 \pmod 3$ while the AHC has $C=-1 \pmod 3$. 
    }
    \label{fig:phase_identification}
\end{figure*}

\PRLsection{Observables}
We now analyze the observed phases of $\lambda$-jellium. Fig.~\ref{fig:phase_identification} shows the real-space charge density $\rho(\v{r})$, the momentum distribution $n(\v{k})$, and the static structure factor $S(\v{q})$ at one representative point in each of the four phases. The liquid is characterized by a uniform density $\rho(\v{r})$, a sharp jump in $n(\v{k})$ at the Fermi level $k_F$, and a smooth structure factor. The three other phases display a localized charge density profile, suggesting crystallization to a triangular lattice commensurate with the system size.
We note that we have not tested for rectangular or rhombohedral orderings, which are disfavored by our choice of simulation cell.
Electron crystallization, i.e. spontaneous breaking of continuous translation and rotation symmetries, is confirmed by the lack of a sharp Fermi surface in $n(\bs{k})$ and the presence of six Bragg peaks in $S(\bs{q})$.

To distinguish the AHC and HWC from the conventional WC
in VMC, we exploit
rotation symmetry. Recall that, at the mean field level, the character of these phases is encoded in the angular momentum eigenvalues~\cite{Turner_2012,fang2012bulk}  of their single particle orbitals at the C$_3$ symmetric points  $\gamma, \kappa, \kappa'$ \footnote{Following the same geometry convention as in Ref.~\cite{AHC3}}.
For C$_3$ symmetry, the Chern number ~\cite{fang2012bulk} satisfies $C = \ell_\gamma + \ell_\kappa + \ell_{\kappa}' \mod 3$, where $\ell_{\gamma,\kappa,\kappa'}$ are the angular momenta at the $\gamma,\,\kappa,\,\kappa'$ points. Further, the change in orbital character means $\ell_\gamma = 0$ for WC and AHC, but $\ell_\gamma = 1$ for the HWC. At the many-body level this information is captured by the eigenvalue of global C$_3$ rotation for different system sizes. 
When the system has $L\times L$ crystal unit cells and $L$ is a multiple of $3$, all C$_3$ symmetric momenta are fully occupied,
and the many body rotation eigenvalue for one of our NQS states $\ket{\Psi}$ is related to the Chern number $C$ by
\begin{equation}
\label{eq:indicator_Chern_number}
    \braket{\Psi|\hat{C}_3|\Psi} = e^{i \frac{2\pi}{3} C},
\end{equation}
where $\hat{C}_3$ rotates the wavefunction counterclockwise by $\frac{2\pi}{3}$ around the crystal's rotation center (see App.~\ref{App:C_3_op}).
Using this approach, the insets in  Fig.~\ref{fig:phase_identification} show 
$\braket{\Psi|\hat{C}_3|\Psi}$, whose 
phases confirm an AHC
with $C=-1~\pmod 3$. 
We can further distinguish the HWC from the WC from the many body angular momentum when $L$ not a multiple of $3$. This is carried out in App.~\ref{App:observables}, where we identify the crystal at small $\lambda$ as the regular WC and the $C=0$ crystal at large $\lambda$ as a HWC with $\braket{\hat{C}_3} = e^{i2\pi/3}$ (see Table~\ref{tb:C3_eigenvalues} for $\hat{C}_3$ expectation values).
This is further evidenced  from the fact that its nontrivial angular momentum at $\gamma$ manifests as a ``halo'' in momentum distribution, with a local minimum at $\v{k}=0$ (Fig.~\ref{fig:phase_identification}).

\PRLsection{Discussion \& Outlook }
This work investigated the effect of quantum geometry on electron crystallization using state-of-the-art neural quantum state VMC calculations. We found that adding Berry curvature and a non-trivial spinor to jellium fundamentally alters its crystallization physics, stabilizing a ``halo" Wigner crystal and a topological anomalous Hall crystal. Moreover, the nature of the crystallization transition is strongly modified from conventional jellium in two crucial ways. 

First, crystallization in the presence of Berry curvature and quantum geometry occurs at much weaker interactions. 
For example, the AHC transition is at $r_s^\star \approx 8$ for $\lambda=0.72$.
To put this in perspective, this means that if quantum geometry was added to a quadratic band without changing the effective mass or dielectric constant, this would increase the critical density for crystallization by as much as $\frac{n^{\star}|_{\lambda=0.72}}{n^{\star}|_{\lambda=0}}=\left(\frac{r_s^{\star}|_{\lambda=0}}{r_s^{\star}|_{\lambda=0.72}}\right)^2\approx 10$ --- about an order of magnitude higher density than jellium.

Second, adding quantum geometry improves the robustness of the crystalline transition both numerically and physically. This manifests clearly in the variational ground state energy $E_0$.
Within our resolution we observe a jump in $\frac{dE_0}{dr_s}$ at the transition, whose magnitude grows significantly when quantum geometry is added (see Fig.~\ref{fig:dEdrs_gr}).
This implies the phase boundaries at finite $\lambda$ to be
vastly more robust against incremental improvement of the variational energies and other numerical errors.
Experimentally, a large slope of the energy difference between liquid and crystal around the transition suggests that the transition region is less susceptible to local inhomogeneities, which can pin short-ranged crystalline correlations and preempt the development of long-range crystalline order~\cite{ge2025visualizing}. However, this expectation depends on additional factors, including the distinct response of each phase to disorder, which may even lead to the opposite trend, a detailed analysis of which will be addressed in future works. 
Lastly, based on the aforementioned reasons, we predict this model to be a promising framework for studying intermediate `micro-emulsion' phases both experimentally and numerically. Such phases were predicted to exist due to an obstruction to first-order transitions in this context~\footnote{In 2d systems of charged particles, interacting through Coulomb or single-gate-screened Coulomb interaction, first order phase transitions as function of density are prohibited~\cite{emery1993frustrated,spivak2003phase,spivak2004phases, jamei2005universal}. Instead, we expect to find a sequence of continuous transitions from FL to WC 
through one or more intermediate phases. There are several theoretical suggestion for the nature of these intermediate phases~\cite{oganesyan2001quantum,spivak2004phases, kim2022interstitial, kim2024dynamical}, with a conclusive numerical determination of its nature currently absent. The range of densities where these intermediate phases appear is expected to increase with increasing chemical potential jump in the putative FL to WC first-order transition. This jump in the chemical potential is related to the  observed jump in $\frac{\partial E_0}{\partial r_s}$ discussed in the text, found to be significantly larger at finite $\lambda$.
Therefore, at finite $\lambda$ we expect these `micro-emulsion' phases to serve as the ground state for a larger range of densities. Lastly, we note that while these phases are unavoidable only in the presence of sufficiently long-range interactions, they may survive even in dual-gated systems
~\cite{sung2025electronic}.}.

Our work paves the way for using NQS to study unexplored electronic ground states of band minima with quantum geometry, where interactions may stablize states such as: generalized flavor-magnetic phases; fractional AHCs, which are electronic crystals composed of fractionally charged anyons;
and exotic chiral superconductors.

\noindent \textit{Acknowledgments}---We thank Steven A. Kivelson, Vladimir Calvera, Yixiao Chen and Conor Smith for help and valuable discussions.
This research was supported in part by grant NSF PHY-2309135 to the Kavli Institute for Theoretical Physics (KITP). Y.Y. acknowledges support from National Science Foundation (NSF) grant No. DMR-2532734 and the use of the Hofstra Star HPC cluster, which was funded by the NSF Major Research Instrumentation (MRI) program (grant No. 2320735).  DEP acknowledges startup funds from UC San Diego. A.V. and J.D. acknowledge funding from NSF DMR2220703 and the Simons Collaboration on Ultra-Quantum Matter, which
is a grant from the Simons Foundation (651440, A.V.).
Y.V. and E.B. were supported by the Simons Foundation Collaboration on New Frontiers in Superconductivity (Grant SFI-MPS-NFS-00006741-03), 
by NSF-BSF Award No. DMR-2310312, and by CRC 183 of the Deutsche Forschungsgemeinschaft (Project C02). 
This research is funded in part by the Gordon and Betty Moore Foundation’s EPiQS Initiative, Grant GBMF8683 to T.S.
The Flatiron Institute is a division of the Simons Foundation. 

%

\clearpage

\appendix
\twocolumngrid

\section{Robustness}
\label{app:numerical_robustness}

\begin{figure}
    \centering
    \includegraphics[width=\linewidth]{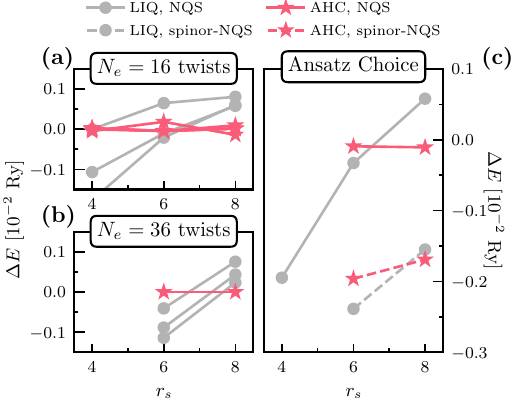}
    \caption{
    \textbf{(a-b)} Relative energy per electron of liquid and AHC states for different twisted boundary conditions at $\lambda=0.72$ for \textbf{(a)} $N_e=16$ and \textbf{(b)} $N_e=36$. Here $\Delta E$ is defined relative to the twist-averaged AHC energy in the $N_e=16$ case, and relative to the un-twisted AHC energy in the $N_e=36$ case. \textbf{(c)} Energies of the liquid and AHC with spinor dependent backflow (spinor-NQS) and with spinor independent backflow (NQS) for $N_e=16$. 
    }
    \label{fig:robustness}
\end{figure}

We now demonstrate that our conclusions --- in particular the existence of the AHC phase ---  are robust to (I) finite size effects (II) using a yet-more-expressive ansatz, and (III) band mixing effects. Due to computational cost, we focus on $\lambda=0.72$ in the vicinity of the crystallization transition, where the energetic competition is particularly sensitive.

\textbf{Finite Size Effects} --- Finite size effects can be grouped into (i) differences arising from a summation of $k$-points on a finite mesh in reciprocal space rather than an integration (quadrature errors) (ii) changes in the correlation functions as a function of the number of electrons \cite{Holzmann_TheoryFiniteSize_2016}.
The dominant one-body contribution to (i) is given by shell effects. These are usually negligible for crystalline phases but can play a significant role in the case of liquid states. 
Shell effects can be quantified and drastically reduced by twist averaging, i.e. averaging over multiple systems with twisted boundary conditions.
We here assess shell effects by employing twisted boundary conditions. Further, we address beyond-shell finite-size effects by confirming the consistency of the liquid-AHC transition within two system sizes ($N_e=16$ and $N_e=36$).

Explicitly, we consider a parallelogram geometry with dimensions $\v{L}_1 = N_1 \v{a}_1$ and $\v{L}_2 = N_2 \v{a}_2$, where $\v{a}_{1,2}$ generate a triangular lattice where we impose twisted boundary conditions for each electron as $\psi(\v{r}+\v{L}_j) = e^{i2\pi \theta_j} \psi(\v{r})$ for some phases $\theta_{j}$ with $j=1,2$.
Twisting the boundary conditions effectively shifts momenta by $k_j \to k_j + \theta_j/L$, allowing us to quantify and address open-shell effects. 
Fig.~\ref{fig:robustness}(a,b) shows the energies of the liquid and AHC states for $N_e=N_1 \times N_2 = 4\times 4 = 16$ and $N_e=6 \times 6 = 36$ electrons. We focus on $\lambda=0.72$ for interaction strengths in the range $r_s = 4-8$, the vicinity of the transition. For each point, we consider three twisted boundary conditions drawn from the Halton quasi-random sequence \cite{halton1960efficiency, qin2016benchmark}.
For $N_e=16$ electrons, shell effects are substantial, 
leading to variation in the crystallization transition: $r_s^\star(N_e=16) \approx 4 - 6.5$. Increasing the system size to $N_e=36$ improves finite size effects substantially (see Fig.~\ref{fig:robustness}(b).
In particular, these energy differences in between different choices of twist are significantly smaller than the energy advantage of the AHC over the liquid at larger $r_s$. While our simulations in the main text do not include twist averaging, these results indicate that the AHC phase should persist in the thermodynamic limit. For $N_e=36$ electrons at $\lambda=0.72$, we estimate $r_s^\star = 7 \pm 0.5$.
We comment further on finite-size effects in the presence and absence of Berry curvature in App.~\ref{App:ansatz}.

\begin{figure}
    \centering
    \includegraphics[width=\linewidth]{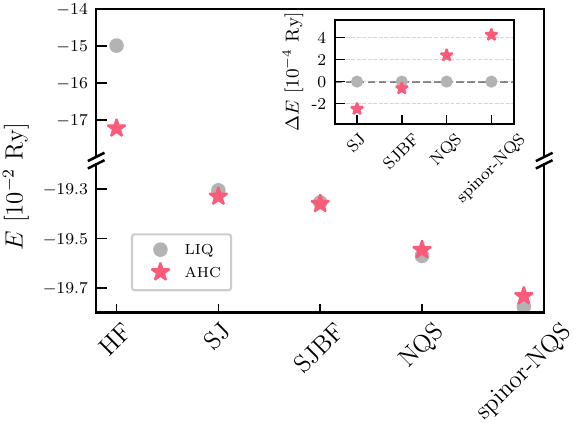}
    \caption{Variational energy per electron for liquid and AHC states at $r_s =6, \lambda=0.72, N_e = 16$, for five variational ans\"atze with increasing expressive power. The ans\"atze, described in the text, are: self-consistent Hartree Fock (HF), Slater-Jastrow (SJ), SJ with backflow (SJBF), SJBF with a spinor-independent neural quantum state (NQS), and SJBF with a spinor-dependent NQS (spinor-NQS). The main text uses the NQS ansatz. Inset: energy differences between the liquid and AHC states.}
    \label{fig:ansatz_dep}
\end{figure}

\textbf{Choice of Variational Ansatz} --- 
Any choice of variatinal wave-function comes with a potential bias. We study this bias by analyzing the competition between liquid and AHC near the transition under a sequence of successively more expressive ans\"atze (see Fig.~\ref{fig:ansatz_dep}).
The simplest ansatz, self-consistent Hartree-Fock (HF), shows competition on the 
$10^{-2}$\si{Ry} level. The next simplest ansatz is Slater-Jastrow (SJ), that has a single Slater determinant multiplied by a two-body Jastrow factor $e^{-U_2(\v{R},S)}$, where $U_2$ is B-spline described below Eq.~\eqref{eq:jastrow}. This relatively simple beyond mean field ansatz pushes the energetic competition down to the $10^{-4}$\si{Ry} level. The next level is the Slater-Jastrow-Backflow (SJBF) ansatz, whose Slater determinant is evaluated at coordinates $\tilde{\v{r}}_i = \v{r}_i + \sum_j B(|\v{r}_i -\v{r}_j|)(\v{r}_i -\v{r}_j)$, where $B$ is parametrized using polynomial B-spline functions.  The neural quantum state (NQS) ansatz, described in the main text, uses neural nets for both the Jastow factor and the backflow function. Its additional expressiveness decreases the optimized energies of both phases on the $10^{-3}$\si{Ry} level (Fig.~\ref{fig:ansatz_dep}). Previous work has shown this ansatz is capable of accurately determining crystallization transitions in the vanilla jellium model~\cite{smith2024ground}.

We now investigate whether a yet-more-expressive ansatz is required to account for the spinor structure of $\lambda$-jellium. A potentially-important shortcoming of the NQS ansatz used in this work is that its backflow function $\mathcal{N}$ depends only on the position of the other electrons, but not their spinor structure. That is, $\tilde{\r}_i = \r_i + \mathcal{N}_i(\v{R})$. We are thus motivated to consider a spinor-NQS ansatz with a spinor-dependent backflow $\tilde{\r}_i = \r_i + \mathcal{N}_i(\v{R},S)$. While more expressive, this ansatz is qualitatively more expensive; the computational cost of evaluating the energy of a spinor-NQS wavefunction is $\mathcal{O}(N)$ more expensive than NQS (see App.~\ref{App:complexity}). Fig.~\ref{fig:ansatz_dep} show it again decreases the energies of both phases substantially over NQS. We speculate this energetic improvement may be from enabling the electrons to follow a local spinor texture. Fig.~\ref{fig:robustness}(c) show this improvement is roughly equal for both the liquid and crystal phases, and that its magnitude is largely independent of $r_s$. Because of this, upgrading from NQS to spinor-NQS changes the transition by $\Delta r_s^\star \approx 1$.  This indicates that the absence of spinor-aware backflow does not substantially bias the NQS ansatz towards liquids or crystals. We thus employ the substantially-cheaper NQS ansatz throughout the main text.

\begin{figure}
    \centering
    \includegraphics[width=\linewidth]{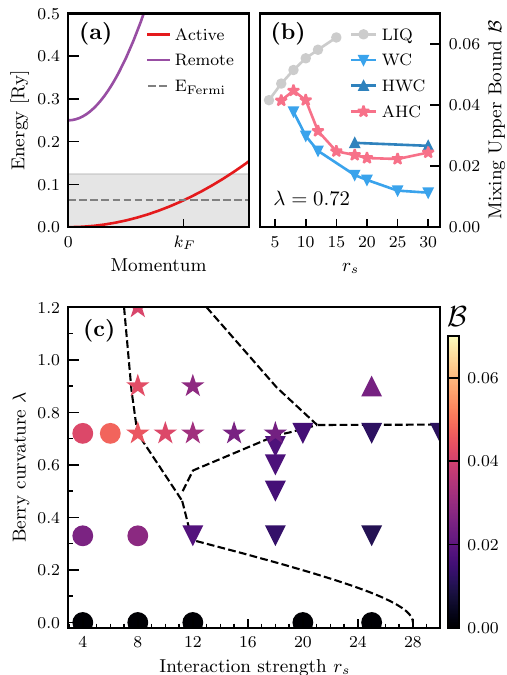}
    \caption{ 
    \textbf{(a)} Dispersion of active and remote bands at $\Delta=2/r_s$ and $r_s=8$.
    The gray region denotes the interaction scale $1/r_s$. \textbf{(b)}    
    Upper bound $\mathcal{B}$ on the upper-band weight of the single-particle density matrix, computed via Eq.~\eqref{eq:upperbound} at $\lambda=0.72$ as a function of $r_s$.
    \textbf{(c)} Map of $\mathcal{B}$ across the phase diagram. Symbols and phase boundaries (dashed lines) match Fig.~\ref{fig:phase_diagram}. One can see that the ``most frustrated" region is at $r_s \lesssim 10$ and intermediate $\lambda$.
    }
    \label{fig:band_mixing}
\end{figure}

\textbf{Quantifying Band Mixing Effects} --- The $\lambda$-jellium model has two bands: the ``active" band at low energies and a remote band whose minimum kinetic energy is set by the adjustable parameter $\Delta$ in Eq.~\eqref{eq:ham}. In the limit $\Delta \to \infty$, one expects the ground state is projected to the Hilbert space of the active band. To limit potential optimization issues associated with the projected limit, we take a finite gap $\Delta = 2/r_s$, i.e. twice the interaction scale of $1/r_s$ (Fig. ~\ref{fig:band_mixing}). Perturbation theory therefore suggests an appreciable amount of band-mixing can occur, whose magnitude we now quantify. 

Consider the term in the Hamiltonian that dictates the spinor structure  
\begin{equation}
     \hat{h}= \begin{bmatrix}
        -\lambda^2 \nabla^2 & i\lambda \partial\\
        i\lambda \bar{\partial} & 1
    \end{bmatrix},
\end{equation}
which is scaled by $\Delta$ in Eq.~\eqref{eq:ham}. Any single-particle wavefunction in the lower band is an exact zero-mode of this operator, whereas the upper band has eigenvalue $\varepsilon^{(\text{upper})}_{\k}=\lambda^2|\k|^2+1$ at momentum $\k$. Single particle orbitals are resolved in the basis of $\hat{h}$ as $\v{\phi}(\v{r}) =\sum_{\k}\left(A_{\k}\v{v}_{\k} +B_{\k}\v{u}_{\k} \right)e^{i\k\cdot\r}$, where $\v{v}_{\k}$ and $\v{u}_{\k}$ are normalized eigen-spinors of the lower and upper band respectively, and $\sum_{\k}|A_{\k}|^2+|B_{\k}|^2=1$. We have $
    \braket{\v{\phi}|\hat{h}|\v{\phi}}=\sum_{\k} |B_{\k}|^2\varepsilon^{(\text{upper})}_{\k}\geq \sum_{\k} |B_{\k}|^2$,
which means that the expectation value $\braket{\v{\phi}|\hat{h}|\v{\phi}}$ is an upper bound on the weight of the single-particle density matrix in the upper band. Thus the VMC ground state expectation
\begin{equation}
    \mathcal{B} := \frac{1}{N_e} \sum_{j=1}^N \braket{\hat{h}_j}
    \label{eq:upperbound}
\end{equation}
is an \textit{upper bound} for the average weight on the upper band. Fig.~\ref{fig:band_mixing}(b) shows the mixing is on the level of a few percent at $\lambda=0.72$, and is higher in the liquid state than the AHC or WC. This indicates band mixing does not bias the state towards crystallization and indeed the liquid may benefit more from band-mixing. Fig.~\ref{fig:band_mixing}(c) shows $\mathcal{B} \to 0$ as $\lambda\to 0$, which is expected, since $[\hat{H},\sigma^z]=0$ there. More broadly, this indicates the active band is ``most frustrated" at moderate $\lambda$ and small $r_s$. We note that this difference in band mixing between the liquid and the crystal is reduced upon introduction of a spin-dependent backflow transformation in the case of $N=16$ electrons. Thus, a potentially imperfect band projection ability of the chosen variational wave-function may introduce a bias against the liquid wave-function.

\section{Energetics of Crystallization}
\label{app:energetics_details}

\begin{figure}
    \centering
    \includegraphics[scale=1]{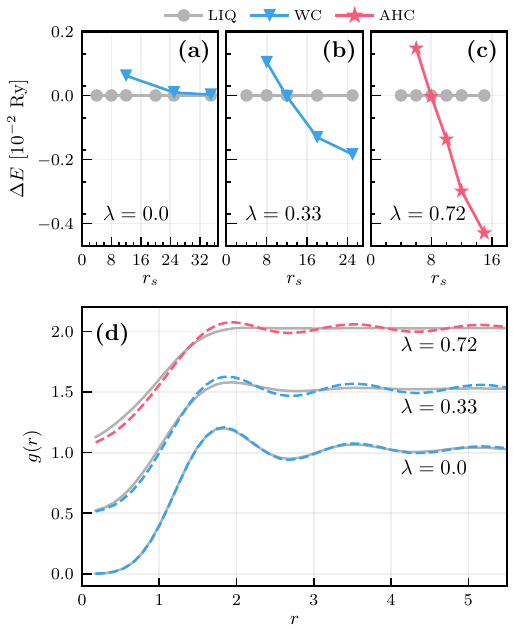}
    \caption{
    \textbf{(a), (b), (c)} The energy difference (per particle) between the VMC crystal and the VMC fluid as a function of $r_s$ for $\lambda=0, 0.33, 0.72$. The magnitude of the slope $|dE/dr_s|$ increases with increasing $\lambda$. \textbf{(d)} The spherically averaged pair correlation function  $g(r)$ (see Eq.~\ref{eq:gofr}) of both the VMC liquid state (grey) as well as the VMC crystal state (blue/red) at $r_s=12$. The $g(r)$ corresponding to the values $\lambda = (0, 0.33, 0.72)$ are plotted with the respective offsets $(0, 0.5, 1)$. We note that $r$ is given in units of $r_s a_B$. All simulations are carried out with $N=36$ electrons.
    }
    \label{fig:dEdrs_gr}
\end{figure}

\begin{figure}
    \centering
    \includegraphics[scale=1]{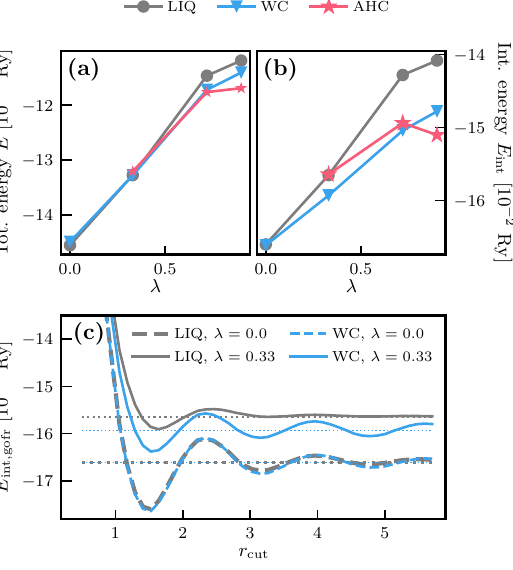}
    \caption{
    \textbf{(a)} The total energies per particle at $r_s=12$ as a function of $\lambda$ for liquid, wigner crystal and AHC states. \textbf{(b)} The interaction energies per particle at $r_s=12$ as a function of $\lambda$ \textbf{(c)} The interaction energy from considering electrons up to distance $r_{\rm cut}$ at $r_s=12$ for $\lambda=0$ and $\lambda=0.33$, as a function of $r_{\rm cut}$ as defined in Eq.~(\ref{eq:egofr}). Here, the horizontal lines refer to the actual interaction energies (corresponding to panel (b) and the limit $r_{\rm cut}\to \infty$. From top to bottom, the horizontal dotted lines refer to the interaction energies of the liquid at $\lambda=0.33$ (grey), the WC at $\lambda=0.33$ (blue), the WC at $\lambda=0$ (blue) and the liquid at $\lambda=0$ (grey). We note that the reference interaction energies at $\lambda=0$ are almost degenerate. Further, $r_{\rm cut}$ is given in units of $r_s a_B$. All simulations are carried out with $N_e=36$ electrons.
    }
    \label{fig:interaction}
\end{figure}

This Appendix investigates the energetic competition near the crystallization transition.

First, we examine the magnitude of the energetic competitions. 
Figure~\ref{fig:dEdrs_gr} demonstrates that the magnitude of the jump in $\frac{dE_0}{dr_s}$ at the transition grows significantly as $\lambda$ increase, where $E_0$ is the ground-state energy (compare the slope of the crystalline states in panels (a), (b), and (c)). Thus, we expect both the liquid and crystal phases to be significantly more stable in numerics than in the standard jellium model.

Second, we proceed to examine the origin of both the shifted crystallization transition as well as the larger energetic competition scale in the presence of quantum geometry. The first hint is given by the spherically averaged pair correlation function $g(r)$, defined in Eq.~\eqref{eq:gofr}, in Fig.~\ref{fig:dEdrs_gr}(d): At $\lambda=0, r_s=12$, $g(r)$ of the liquid and the crystal are indistinguishable by eye, hinting to energetically close interaction energies. 
As $\lambda$ increases, their behaviors start to differ: The (short-range) oscillations in general get suppressed and the value of $g(r=0)$ (which is linked to the exchange-correlation hole) increases. Both of these effects are larger for the liquid. Below, we quantify this behavior and the role of $g(r)$ further by directly examining the interaction energies.

Fig.~\ref{fig:interaction}(a,b) show the total energies and the interaction energies at $r_s=12$ as a function of $\lambda$ for liquid, Wigner crystal and AHC ans\"{a}tze. We find that both the total energy and the interaction energy for all states --- independent of their nature --- increase with increasing $\lambda$. However, we find the interaction energy increases more quickly for liquids, i.e. quantum geometry penalizes the liquid more strongly. This effect is incompletely canceled out by the kinetic energy, ultimate causing a decrease in $r_s^\star$ as $\lambda$ increases.

In Fig.~\ref{fig:interaction} (c), we show that 
dominant contribution to the change in the interaction energy is due to to the short-range part of the interaction, i.e. the exchange-correlation hole. To quantify this, we define 
\begin{align}
E_{\rm int}(r_{\rm cut}) =  \frac{\pi}{A}\int_0^{r_{\rm cut}} r dr \, \left[g(r)-1\right]v(r),
\label{eq:egofr}
\end{align}
where $A$ the area of the simulation cell. Physically, $E_{\rm int}(r_{\rm cut})$ measures the contribution to the interaction energy per electron that comes from all particles that have less than distance $r_{\rm cut}$ to each other. Thus, by plotting 
$E_{\rm int}(r_{\rm cut})$
as a function of the upper integration bound $r_{\rm cut}$ and comparing to the reference interaction energies $\lim_{r_{\rm cut}\to \infty}E_{\rm int}(r_{\rm{cut}}) =E_{\rm int}$ (plotted as dotted horizontal lines for reference), we are able to quantify whether short-range or long-range correlations dominate the interaction energy. Figure~\ref{fig:interaction} shows that all changes in the interaction energy 
already manifest at small values of the integration cutoff, $r_{\rm cut} \approx r_s$. This includes the energetic differences between the states at $\lambda=0$ and $\lambda=0.33$ (dashed and solid lines, respectively), as well as the increasing interaction energy differences between the liquid and the crystal (gray and blue lines). These data indicate that the effect of quantum geometry on the exchange-correlation hole plays a main part in the shift of the critical $r_s$.

We speculate that the presence of quantum geometry hinders an energy-efficient creation of a exchange-correlation hole --- an effect potentially more relevant in liquid states, where electrons are not already kept apart via a crystalline structure. This intuition is compatible with our findings, that a liquid ground state is disfavored upon increasing $\lambda$.

\section{Phase Identification from Symmetry Indicators}
\label{App:C_3_op}
In order to identify the different crystalline phases, we must calculate the expectation value of $\hat{C}_3$, the operator that counter-clockwise by $\frac{2\pi}{3}$ about the crystal's rotation center. We first need to find a good rotation center $\r_0$, for which the shifted wavefunction $\tilde{\Psi}(\R,S)=\Psi(\R-\R_0,S)$, where $\R_0\equiv(\r_0,\cdots,\r_0)$, is symmetric under rotations around its origin (i.e. $\tilde{\Psi}$ is an eigenvector of $\hat{R}_{2\pi/3}$). We identify the rotation center by folding $\rho(\r)$ into the unit cell of the symmetry broken crystal, and take the position of its maximum
\begin{equation}
    \r_0=\arg\max_{\r}\left[ \sum_{\v{a}}\rho(\r+\v{a})\right],
\end{equation}
where $\v{a} = n_1 \v{a}_1 + n_2\v{a}_2$ runs over direct lattice vectors, $\r$ is a point in the fundamental unit cell, and $\rho(\r)\equiv\braket{\hat{\rho}(\r)}$ is the expectation value of the density operator.
We then calculate $\braket{\Psi|\hat{C}_3|\Psi}\equiv\braket{\tilde{\Psi}|\hat{R}_{2\pi/3}|\tilde{\Psi}}$ using Monte-Carlo sampling.
$\hat{R}_{2\pi/3}$ acts on the wavefunction as
\begin{equation}
    \bra{\R,S}\hat{R}_{2\pi/3}\ket{\Psi}=\Psi\left(\{\mathcal{R}_{-\frac{2\pi}{3}}\r_i\},S\right) \prod_{j=1}^{N_e} e^{i\frac{2\pi}{3}\frac{\sigma_j+1}{2}},
\end{equation}
where $\mathcal{R}_{\theta}$ is the $2 \times 2$ matrix acting on a single particle position $\r_i$, corresponding to counter-clockwise rotation by $\theta$, and $\{\mathcal{R}_{-\frac{2\pi}{3}}\r_i\}=\left(\mathcal{R}_{-\frac{2\pi}{3}}\r_1,\cdots,\mathcal{R}_{-\frac{2\pi}{3}}\r_{N_e}\right)$.

\begin{table}
\centering
\begin{tabular}{ p{0.5in} p{0.4in}  p{0.4in} c c}
\toprule
Phase & $r_s$ & $\lambda$ & \textbf{$L=4$} & \textbf{$L=6$} \\
\midrule
WC & $25$ & $0.33$ & $0.986\cdot e^{0.0002i}$ &  $1.032\cdot e^{0.001i}$\\
HWC & $25$ & $0.9$ &  \colorbox{gray!30}{$0.9637\cdot e^{\frac{1.94\pi}{3}i}$} &  $0.999\cdot e^{-0.0001i}$\\
AHC & $12$ & $0.72$ & $0.993\cdot e^{0.001i}$ &  \colorbox{gray!30}{$0.997\cdot e^{-\frac{1.99\pi}{3}i}$}\\
\bottomrule
\end{tabular}
\caption{A table of $\braket{\hat{C}_3}$ calculated within NQS-VMC for each of the phases, for system sizes of $L\times L$ crystal unit cells, with $L=4,6$. Gray boxes indicate a non-trivial $\hat{C}_3$ expectation value.}
\label{tb:C3_eigenvalues}
\end{table}

To validate our choice of rotation center, we
use the fact that $\hat{C}_3$ is a bounded operator satisfying $|\braket{\hat{C}_3}|\leq1$, and $|\braket{\Psi|\hat{C}_3|\Psi}|=1$ if and only if $\ket{\tilde{\Psi}}$ is an eigenvector of $\hat{C}_3$. Therefore $|\braket{\Psi|\hat{C}_3|\Psi}|$ is a good measure for the extent to which the wavefunction is indeed $C_3$ symmetric around $\r_0$. For all crystals we find $|\braket{\Psi|\hat{C}_3|\Psi}|\approx1$,
confirming the validity of our rotation centers.

As remarked in the text, for systems of size $L\times L$ where $L\mod 3 = 0$, $\braket{\hat{C}_3}$ allow us to detect the Chern number $C (\text{mod } 3)$, and therefore distinguish AHC from both WC and HWC. Whereas for systems where $L\mod 3 \ne 0$, the expectation value $\braket{\hat{C}_3}$ further distinguishes between WC and HWC. In our case, we consider systems of size $6 \times 6$ and $4\times 4$ at specific points within each of the crystalline phases. We classify those phases based on the expectation value of $\braket{\hat{C}_3}$ (assuming both system sizes converged to the same phase)
These findings are summarized in Table~\ref{tb:C3_eigenvalues}.

\section{Estimating Observables Through Monte Carlo Sampling}
\label{App:observables}

This Appendix briefly reviews the definitions of the various observables used in the main text and the method by which they are calculated from the NQS wavefunction.

\subsection{Monte Carlo Sampling for Local Observables}
Consider a non-normalized many-body wavefunction $\ket{\Psi}$ represented numerically by access to its wavefunction amplitudes $\Psi\left( \R,S\right)$ such that
\begin{equation}
    \ket{\Psi}=\sum_S\int d\R\Psi\left( \R,S\right)\ket{\R,S},
\end{equation}
with $\v{R}=(\r_1,\cdots,\r_N)$ the many-body spatial coordinate and $S=(\sigma_1,\cdots,\sigma_N)$ the many-body spinor coordinate. 
Our goal is to evaluate the expectation value of a generic many-body operator $\hat{\mathcal{O}}$ through Monte-Carlo sampling of coordinates $(\R,S)$.

The expectation value is written as
\begin{align}&\braket{\hat{\mathcal{O}}}=\frac{\braket{\Psi|\hat{\mathcal{O}}|\Psi}}{\braket{\Psi|\Psi}} \\
    & = \sum_{S,S'}\int d\R d\R'\; \frac{\Psi^*(\R,S)\braket{\R,S|\hat{\mathcal{O}}|\R',S'}\Psi(\R',S')}{\braket{\Psi|\Psi}} \\
     &\equiv \sum_{S}\int d\R f\left( \R,S \right)\mathcal{O}_{loc}\left( \R,S \right),
\end{align}
where in the last equation we defined a probability density $f$ and $\mathcal{O}_{loc}$, a local estimator of $\hat{\mathcal{O}}$ as 
\begin{align}
    &f\left( \R,S \right)=\frac{\left|\Psi\left( \R,S \right)\right|^2}{\sum_{S}\int d\R\left|\Psi\left( \R,S \right)\right|^2}, \\   &\mathcal{O}_{loc}\left( \R,S \right) =\frac{\sum_{S'}\int d\R' \braket{\R,S|\hat{\mathcal{O}}|\R',S'}\Psi\left( \R',S' \right)}{\Psi\left( \R,S \right)}.
\end{align}
While $f$ is a proper probability density, sampling it directly is hard since it requires calculating $\sum_{S}\int d\R\left|\Psi\left( \R,S \right)\right|^2$ first. Instead we estimate can estimate any integral $\sum_{S}\int d\R f\left( \R,S \right)[\cdots]$ through a Monte-Carlo sampling of coordinates $\R,S$ with update rules that satisfy detailed balance, using the easily accessible ratio $\frac{f\left( \R',S' \right)}{f\left( \R,S \right)}=\frac{\left|\Psi\left( \R',S' \right)\right|^2}{\left|\Psi\left( \R,S \right)\right|^2}$. 
For operators $\hat{\mathcal{O}}$ that are local in the spatial coordinate $\hat{\mathcal{O}}\Psi\left( \R,S\right)\ket{\R,S} = \sum_{S'} \ket{\R,S'}\mathcal{O}_{S',S}(\R)\Psi\left( \R,S\right)$, we can simplify it to
\begin{equation}
    \mathcal{O}_{loc}\left( \R,S \right) =\frac{\sum_{S'}\mathcal{O}_{S,S'}(\R)\Psi\left( \R,S' \right)}{\Psi\left( \R,S \right)}.
    \label{eq:local operator}
\end{equation}
All the observables we calculate are either local or bi-local in the many-body spatial coordinate and sparse in the spinor degree of freedom (i.e. $\mathcal{O}_{S,S'}$ is non-zero for a set $\{S'_i\}$ of polynomial size in the number of electrons). Therefore, for a given $\R,S$ we calculate $\mathcal{O}_{loc}\left( \R,S \right)$ exactly and efficiently. 

\subsection{List of Calculated Observables}
\subsubsection{Energy $E$} 
The energy is given by $E=\braket{\hat{H}}$ with the Hamiltonian $\hat{H}$ defined in Eq.~\eqref{eq:ham}. 
The local energy is given according to Eq.~\eqref{eq:local operator} by
\begin{align}
    &E_{loc}\left( \R,S \right) = V(\R) +\sum_{i=1}^N\frac{\sum_{\sigma'=\pm1}[\hat{h}_i]_{\sigma_i,\sigma'}\Psi\left( \R,S_i' \right)}{\Psi\left( \R,S \right)}, \\
    &\hat{h}_i =\begin{bmatrix}
        -(\lambda^2+\frac{1}{r_s^2}) \nabla_i^2 & i\lambda \partial_i\\
        i\lambda \bar{\partial}_i & 1-\frac{\nabla_i^2}{r_s^2}
    \end{bmatrix}
    \label{eq:E_loc}
\end{align}
where $S=(\sigma_1\cdots\sigma_N)$, $S_i'=(\sigma_1\cdots\sigma_{i-1},\sigma',\sigma_{i+1}\cdots\sigma_N)$ (which is equivalent to $S$ except for the $i$th spinor), and $\hat{h}_i$ is the single particle kinetic term for the $i$th particle.

\subsubsection{Density $\rho(\r)$} 
The density operator is given by $\hat{\rho}(\r)\ket{\R,S}=\sum_{i=1}^N\delta(\r_i-\r)\ket{\R,S}$, with $\R=(\r_1,\cdots,\r_N)$. 

\subsubsection{Momentum distribution $n(\k)$} 

We define
\begin{equation}
    n(\k) = \sum_{\sigma}\braket{c^\dagger_{\k,\sigma}c_{\k, \sigma}}.
\end{equation}
To compute this, we introduce its Fourier transform:
\begin{equation}
n(\k) = \frac{1}{A} \int d{\r}e^{-i{\k}\cdot {\r}} \braket{\tilde{n}({\r})},
\end{equation}
where
\begin{equation}
    \tilde{n}({\r}) = \sum_{\sigma}\int d{\v r}'c^\dagger_{{\v r}'+\v{r},\sigma}c_{\v{r}', \sigma}.
\end{equation}
and $A$ is the area of the simulation cell.
Its action on the wavefunction is given by
\begin{equation}
    \bra{{\bf R},S}\tilde{n}({\bf r})\ket{\Psi}=\sum_{i}\Psi({\bf R}
    : \v{r}_i-\v{r}, S)
\end{equation}
where ${\bf R}:{\v r}_i - {\v r}$ denotes $\left({\v r}_1, ...{\v r}_{i-1}, {\v r}_i - {\v r}, {\v r}_{i+1}, ... {\v r}_N\right)$.

\subsubsection{Pair correlation function $g(r)$}
We define the spherically averaged pair correlation function as~\cite{Giuliani_Vignale_2005}
\begin{align}
 g(r)= \frac{A}{2 \pi r N^2 }\left\langle \sum_{i \neq j} \delta(|{\v r}_i - {\v r}_j|-r)\right\rangle.
 \label{eq:gofr}
\end{align}
Here, $A$ is the volume of the simulation cell.  We estimate $g(r)$ by binning the inter-particle distances $|{\v r}_i-{\v r}_j|$ into histograms.

\subsubsection{Structure factor $S(\v{q})$} 
The static structure factor is given by
\begin{align}
S({\v q}) = \frac{1}{N}\langle \hat{n}_{-{\v q}}\hat{n}_{\v q}\rangle,
\end{align}
where
\begin{align}
\hat{n}_{\v q} = \sum_i e^{-i {\v q}\cdot{\hat{\v r}}_i}
\end{align}
is the Fourier transform of the density operator. Thus, the static structure factor is a measure of the average density fluctuations at wave vector ${\v q}$ \cite{Giuliani_Vignale_2005}.

\section{Neural-network ansatz and optimization details}
\label{App:ansatz}

\subsection{Ansatz}
In our neural-network VMC, 
we parametrize the ground state using
(MP)$^2$NQS~\cite{smith2024ground}, a Slater-Jastrow-backflow (SJB) ansatz that represents electrons with an internal spinor structure
\begin{equation}
    \Psi\left( \R,S \right) = e^{-U\left( \R,S \right)} \det\left[ \phi_{j} \left( \bs{r}_i+\mathcal{N}_i(\R),\sigma_i \right) \right],
\end{equation}
where: $\R=(\r_1,\dots,\r_N)$ is the many-body coordinate, $S=(\sigma_1,\dots,\sigma_N)$ the many-body spinor index in $\hat{z}$ basis ($\sigma_i=\pm1$), $U\left( \R,S \right)$ a permutation invariant `potential' from which the Jastrow factor is derived, $\phi_j$ a set of single particle orbitals, and $\mathcal{N}_i(\R)$ a backflow transformation encoded through a neural-network.

\subsubsection{Orbitals}
The orbitals $\phi_{j}$ are parametrized as a sum of plane-waves basis orbitals, with
\begin{equation}
    \phi_j(\r,\sigma)=\sum_\ell A_{j,\ell}^{(\sigma)} \exp{(i\k_\ell\cdot\r)}~,
\end{equation}
with $\{\k_\ell\}$ being all valid momenta for the system cell with periodic boundary conditions up to some momentum cutoff much larger than the relevant Fermi vector. For the liquid phase the coefficients $A_{j,\ell}$ are not constrained to respect any symmetry, as we find this to give the lowest energy.
For the crystalline phases, we impose discrete translation invariance on the orbitals by 
\begin{equation}
\phi_j(\r,\sigma)= \sum_{\ell}A_{j,\ell}^{(\sigma)} \exp{\left[i(\v{p}_j+\bf{G}_\ell)\cdot\r\right]},
\end{equation} 
where $\v{p}_j$ is a momentum in the first Brillouin-zone of the crystal structure dictated by the imposed translational invariance, and ${\bf{G}_\ell}$ is the set of reciprocal lattice vector of the crystal up to some momentum cutoff. A restriction to commensurate filling $\nu=1$ of the crystal guarantees a unique matching between orbitals $\phi_j$ and momenta $\v{p}_j$ in the first crystal Brillouin-zone.
We choose the simulation cell to be commensurate with a triangular lattice.

\subsubsection{Backflow}
Our neural network backflow is based on a message-passing network. Details of the implementation are given in~\cite{smith2024ground}. In our calculation we set $d_1$, the size of the one body stream, and $d_2$, the size of the two body stream to be $d_1=32$ and $d_2=26$. We use three message-passing layers. For the normal NQS ansatz used throughout the paper we pass $s_{ij}=0$ into $v_{ij}$ the two-body ``visible" feature vector, to keep the backflow independent of the spinor index.
For the spinor-NQS ansatz, used in App.~\ref{app:numerical_robustness}, we pass $v_i=\sigma_i$ as the single-body ``visible" feature instead of a null vector, and we pass $s_{ij}=\sigma_i+\sigma_j$ into the two-body ``visible" feature vector $v_{ij}$.

\subsubsection{Jastrow}
For the Jastrow factor, $e^{-U\left( \R,S \right)}$, we use two terms 
\begin{equation}
    U\left( \R,S \right) = U_2\left( \R,S \right) + U_{N.N.}( \R).
\end{equation}
The first term, $U_2$, is a two-body spinor-dependent Jastrow: $U_2=\sum_{i\neq j}u_{\sigma_i,\sigma_j}(|\r_i-\r_j|)$, where each $u_{\sigma,\sigma'}$ is encoded as a cubic B-spline with $n=9$ segments. 

The additional term, $U_{N.N.}$, is a sum of terms derived by applying a MLP to $h_i$, the single body `hidden' features from the last message passing layer of the backflow transformation, and to the backflow transformed coordinates $\tilde{\r}_i=\r_i+\mathcal{N}_i(\R)$. We use $L=4$ ($3$ hidden layers), with the width of the hidden layers taken to be $32$. The details of the transformation are similar to the transformation used in~\cite{smith2024ground}, with one difference. The difference is that each MLP has two outputs corresponding to the real and imaginary part of $U_{N.N.}$. This allows $U_{N.N.}$ to be complex, which is found to be energetically relevant due to the absence of time reversal symmetry in the model.

\subsection{Optimization}
We optimize the ansatz described above via minimum-step Stochastic Reconfiguration (minSR) \cite{sorella1998green,sorella2001generalized,chen2024empowering}. While the NN-ansatz introduced above is sufficiently expressive to describe any of the here studied phases, we find that random initialization and subsequent optimization to the state with lowest energy is here very rarely stable. Instead, we initialize the variational wave-function within one phase and then proceed to optimize the full wave-function. With very few exceptions, by choosing a sufficiently small learning rate the wave-function typically converges within the chosen initial phase. We verify the nature of the final state by computing physical observables. For each value of $r_s$ and $\lambda$, we then obtain a set of energies associated to wave-functions of different nature (liquid, AHC, WC or HWC). We here detail how the initializations are obtained for the respective phases. In general, we use a combination of mean-field based orbital initialization and transfer learning. 

\subsubsection{Liquid}
In order to obtain an optimized wave-function of liquid nature, we initialize the orbitals by filling the lower band of the non-interacting part of the Hamiltonian in Eq.~(\ref{eq:ham}) within our ${\bf k}$-grid. We find it to be most stable to use this initialization at low $r_s$ (concretely, $r_s=4$) and perform a full first optimization there. Then, once a stable solution is obtained, we use it to transfer learn to larger values of $r_s$ (i.e. initializing all wave-function parameters with the solution at lower $r_s$). 

\subsubsection{WC, AHC, HWC}
We initialize the crystalline orbitals using converged Hartree-Fock solutions within the respective phase. Then, we train a wave-function with Bspline, Jastrow, and Backflow (and allow the orbitals to optimize), which consitutes a simple starting point. The resulting orbitals are then, together with the trained Jastrow factor, used as initializion for the full NN ansatz. Once a fully trained neural network is obtained within the respective phase, we use its parameters as an initialization for different values of $(r_s, \lambda)$, i.e. (step-wise) transfer learning through the phase diagram. The Chern number as well as other observables listed in App.~\ref{App:observables} are then computed to determine the nature of the final state. Typically, the state stays within the phase the orbitals have been initialized in.

\subsubsection{Computational cost of the optimization}
Given a set of initial orbitals, one full neural-network wave-function optimization for $N=36$ electrons takes approximately $24$ hours on four \verb|h200 nvidia| GPUs.

\section{Phase boundaries}
\label{app:boundaries}

\begin{figure}[t]
    \centering
    \includegraphics[scale=1]{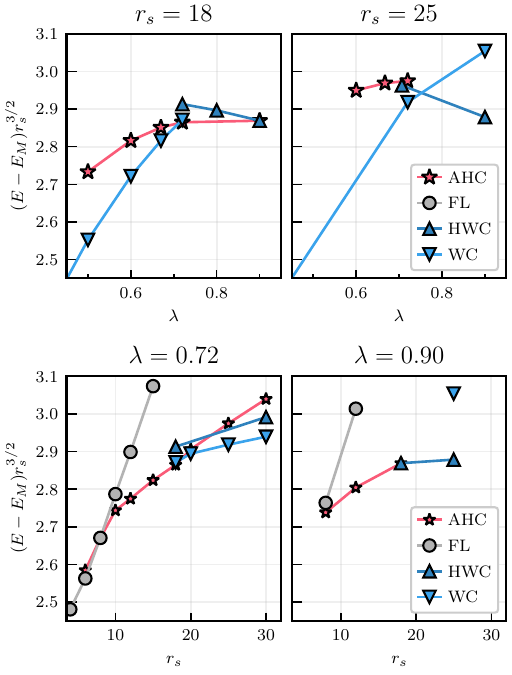}
    \caption{Variational energies per particle for different line-cuts at fixed $r_s$ (top row) and fixed $\lambda$ (bottom row). Here, we subtract the Madelung energy of a triangular crystal $E_M = 2\cdot 1.106103/r_s$ \si{Ry} and rescale the result by $r_s^{3/2}$ in order to make the energetic competition and crossings visible.
    }
    \label{fig:E9_linecuts}
\end{figure}

We perform simulations, on a finite (irregular) grid and estimate the phase boundaries depicted in Fig.~\ref{fig:phase_diagram} via linear piecewise interpolation of energy line cuts. This yields phase transition points at values of $r_s, \lambda$ that lie on the grid lines (the exception is $\lambda=0$, see below). We then connect these points via straight lines to obtain  Fig.~\ref{fig:phase_diagram}.

The top row of Figure ~\ref{fig:E9_linecuts} depicts the energetic competitions as a function of $\lambda$ at fixed $r_s=18,25$. In particular, as $\lambda$ increases, the standard WC is energetically penalized. Instead, the topological crystals are most favorable: the AHC at an intermediate region of $\lambda$ and the HWC at large $\lambda$.

The Bottom row of Figure~\ref{fig:E9_linecuts} shows energy line cuts at fixed $\lambda=0.72$ and $\lambda=0.9$. Both show a region of stability of the AHC, that is bounded by a liquid at low $r_s$. At $\lambda=0.72$ we find a transition to a standard WC at larger $r_s$, but for $\lambda=0.9$, the AHC transitions to a HWC at $r_s^\star \approx 18$.

\subsection{$\lambda=0$}
The scale of the energy differences of competing phases determines the degree of accuracy necessary to resolve the phase boundaries. 
In particular, within standard jellium the energy differences are of order ${\sim}10^{-4}$\si{Ry}, which makes estimations of phase transitions highly sensitive to finite-size effects and other sources of uncertainty --- infamously causing estimates of critical $r_s^\star$ to shift drastically within small improvements of numerics.

\begin{figure}[tb]
    \centering
    \includegraphics[scale=1]{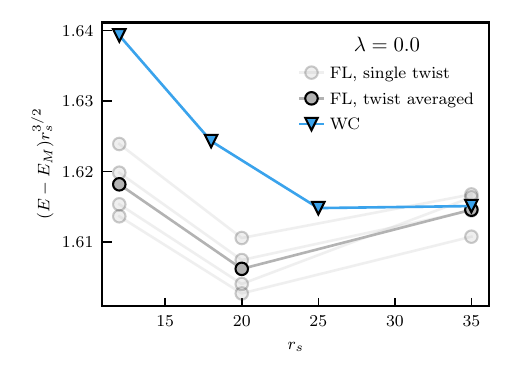}
    \caption{The (rescaled) variational energies per particle of the liquid and WC state at $\lambda=0$. Here, the transparent grey curves denote different twists drawn from the quasi-random Halton sequence~\cite{halton1960efficiency}. The darker grey curve corresponds to the twist-averaged energies of the liquid state.
    }
    \label{fig:lambda0_twists}
\end{figure}

Our estimate of the liquid-crystal transition in polarized jellium is in the range $25\lesssim r_s \lesssim 35$. Even when employing twist-averaging we are not able to pinpoint the transition further without additional simulations or finite-size scaling. More concretely, we estimate the transition by initializing a liquid wave-function (transfer learning from low $r_s$) and a crystal wave-function, and comparing the energies. In addition, we reduce finite-size effects by twist averaging. This procedure yields an apparent transition around $r_s^\star \approx 35$ (see Fig.~\ref{fig:lambda0_twists}).

\begin{figure}[tb]
    \centering
    \includegraphics[width=0.5\textwidth]{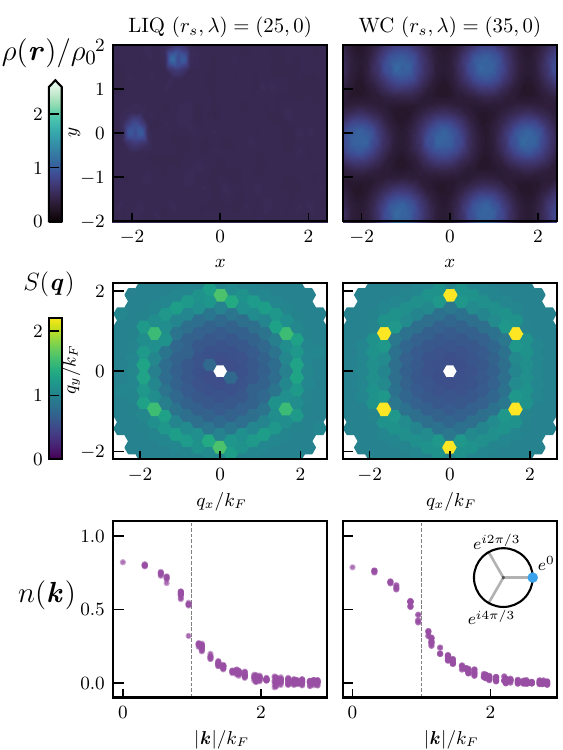}
    \caption{The charge density, structure factor and momentum distribution for the liquid and crystal states close to the phase boundary at $\lambda=0$.
    }
    \label{fig:lambda0_observables}
\end{figure}

However, a subtlety arises requiring closer examination of the potential phase boundaries. Despite being initialized in a liquid phase, the ground state starts accumulating crystalline features starting at $r_s=25$ (see Fig.~\ref{fig:lambda0_observables}). Typically, such a region with unclear ground-state features decreases as the system size is increased, suggesting the origin of this behavior is finite-size effects. In addition, it is possible that transfer learning with smaller step size could yield a sharper transition. As $\lambda=0$ was not the focus of our study, we here refrain from additional simulations due to their computational cost.

\begin{figure}[tb]
    \centering
    \includegraphics[width=0.5\textwidth]{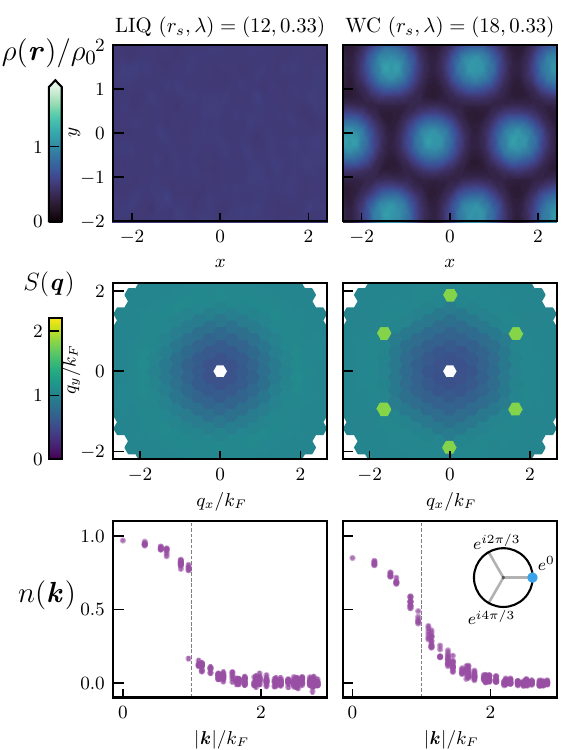}
    \caption{The charge density, structure factor and momentum distribution for the liquid and crystal states at the two closest simulated points on both sides of the liquid-WC phase boundary at $\lambda=0.33$.
    }
    \label{fig:lambda0p33_observables}
\end{figure}

\begin{figure}[tb]
    \centering
    \includegraphics[width=0.5\textwidth]{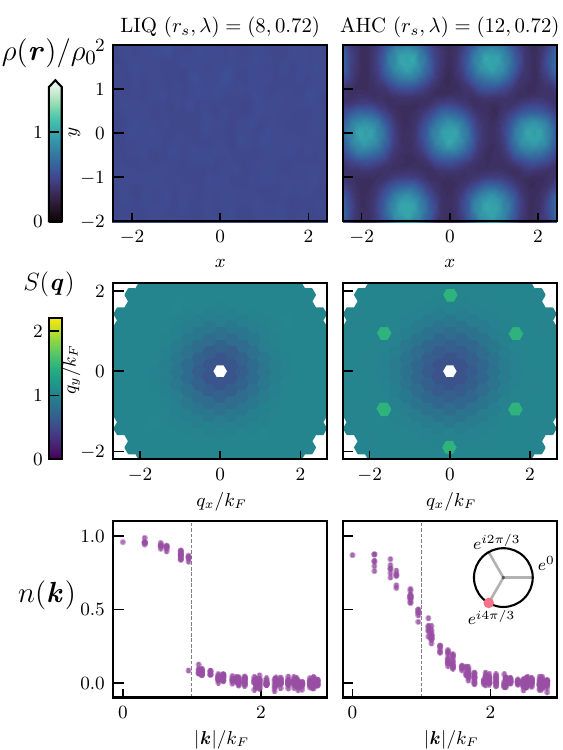}
    \caption{The charge density, structure factor and momentum distribution for the liquid and crystal states at the two closest simulated points on both sides of the liquid-AHC phase boundary at $\lambda=0.72$.
    }
    \label{fig:lambda0p72_observables}
\end{figure}

Instead, we note that the same ambiguities {\it do not appear} at finite $\lambda$. Conveniently, the energy difference scale increases by an order of magnitue as $\lambda$ increases for all finite values of $\lambda$ we have simulated. Thus, estimates of critical $r_s$ away from the $\lambda=0$ line are less susceptible to sources of uncertainty such as finite-size effects, as we showed in App.~\ref{app:numerical_robustness}. To evidence this, Figs.~\ref{fig:lambda0p33_observables} and~\ref{fig:lambda0p72_observables} show the observables of the NQS states near the transition at $\lambda=0.33$ and $\lambda=0.72$, unambiguously identifying their phase character. This allows us to make a straightforward estimate of the phase boundaries when $\lambda>0$.

\section{Complexity of NQS \& Spinor-NQS}
\label{App:complexity}
Consider the local energy $E_{loc}(\R,S)$, given in Eq.~\ref{eq:E_loc}. It requires calculating the WF amplitude $\Psi(\R,S')$ (and its derivatives), for all many-body spinor vectors $S'$ that differ from $S$ by a single-body spinor. Naively, the cost of this operation is $N_e\times C(\Psi)$, with $C(\Psi)$ being the computational complexity of calculating $\Psi$ and its relevant derivatives, and $N_e$ the number of electrons. Within the NQS ansatz, the asymptotic complexity ($N_e\rightarrow\infty$) is dominated by the determinant part. However, for practical system sizes it is often the message-passing graph neural-network (GNN) $\mathcal{N}_i$ that dominates the runtime. For the spinor-independent NQS, the backflow transformation $\mathcal{N}_i(\R)$ is independent of $S$ and can be computed once per every call to $E_{loc}(\R,S)$. As a result, changing $S$ by a single-body spinor modifies only a single-row in the Slater determinant. Such a low-rank update of a determinant can be done efficiently. Therefore, both in the scaling limit ($N_e\rightarrow\infty$) and for realistic $N_e$, the complexity of calculating $E_{loc}(\R,S)$ is reduced to $C(\Psi)$, the complexity of a single call to $\Psi$ (with its relevant derivatives).

For spinor-NQS the backflow $\mathcal{N}_i(\R,S)$ depends explicitly on $S$, therefore such a reduction in the complexity is not possible. The complexity $E_{loc}(\R,S)$ in this case is in fact $N_e\times C(\Psi)$ --- parametrically larger than for (spinor-independent) NQS.

\FloatBarrier
\end{document}